\documentclass[preprint,pteplogo]{ptephy}

\preprintnumber{YITP-14-72} 

\usepackage{amsmath}
\usepackage{graphicx}

\newcommand\sout{\bgroup \color{red} \ULdepth=-.5ex \ULset}

\begin{document}

\title{Momentum scale dependence  of the net quark number fluctuations near chiral crossover}

\author{\name{Kenji Morita}{1,2} and \name{Krzsyztof Redlich}{2,3}}

\address{\affil{1}{Yukawa Institute for Theoretical Physics, Kyoto University,
Kyoto 606-8502, Japan}
\affil{2}{Institute of Theoretical Physics, University of Wroclaw,
PL-50204 Wroc\l aw, Poland}
\affil{3}{Extreme Matter Institute EMMI, GSI, Planckstr. 1, D-64291 Darmstadt, Germany}
\email{kmorita@yukawa.kyoto-u.ac.jp}}

\begin{abstract}%
 We investigate properties of the net baryon  number fluctuations near
 chiral crossover  in a hot and dense medium of strongly interacting
 quarks. The chirally invariant quark-antiquark  interactions are
 modeled by  an effective quark-meson Lagrangian. To preserve remnants
 of criticality in the $O(4)$ universality class,
 we apply the functional renormalization  group method to describe
 thermodynamics near chiral crossover.
 Our studies are focused on the influence of the momentum cuts on the
 critical behavior of  different  cumulants of the net quark number
 fluctuations. We use the momentum scale dependence of the flow equation
 to examine how the suppression of the momentum modes in the infrared
 and ultraviolet regime modifies  generic properties of fluctuations
 expected in the $O(4)$ universality class. We show, that the pion mass
 $m_\pi$ is a natural soft momentum scale at which cumulants are
 saturated at their critical values, whereas  for  scales larger than
 $2m_\pi$  the characteristic $O(4)$ structure of the higher order
 cumulants get lost. These results indicate, that when measuring
 fluctuations of the net baryon number in heavy ion collisions to search
 for a  partial restoration of chiral symmetry or critical point, a
 special care have to be made when introducing kinematical cuts on the
 fluctuation measurements.
\end{abstract}


\maketitle


\section{Introduction}
Exploring possible evidence of partial restoration of chiral symmetry
in a medium crated in heavy ion collisions is one of the most important
and challenging problems \cite{friman11:_cbm_physic_book,fukushima11:_phase_diagr_of_dense_qcd,fukushima13:_phase_diagr_of_nuclear_and}.
Experimental studies along this line have been carried out by measuring
fluctuations of conserved charges, in particular,  of the net baryon number
\cite{aggarwal10:_higher_momen_of_net_proton,STAR_pn_2013} and the
electric charge \cite{adamczyk14:_beam_energ_depen_of_momen}.

Fluctuations of conserved charges are particularly interesting probes of
critical phenomena and phase diagram in QCD.
gThe charge currents couple to the soft ``sigma'' modes, thus
correlations and fluctuations of charge densities are directly affected
by the chiral symmetry restoration at finite temperature and net baryon
number density
\cite{hatta03:_proton_number_fluct_as_signal,stephanov98:_signat_of_tricr_point_in_qcd,ejiri06:_hadron_fluct_at_qcd_phase_trans,karsch11:_probin_freez_out_condit_in,stephanov09:_non_gauss_fluct_near_qcd_critic_point,stephanov11:_sign_of_kurtos_near_qcd_critic_point}. 

For massless two-flavor quarks, the QCD phase transition was
conjectured to be of the second order, and belonging to
the $O(4)$ universality class \cite{pisarski84:_remar_chiral_trans_in_chrom}.
Current lattice QCD (LQCD) simulations at physical quark masses show,
that at vanishing and small baryon density the transition from a hadron
gas to a quark gluon plasma  is crossover \cite{aoki06}. In addition,
LQCD also indicates, that the chiral crossover appears in the critical
region of the second order transition belonging to the $O(2)/O(4)$
universality class
\cite{ejiri09:_magnet_equat_of_state_in_flavor_qcd,kaczmarek11:_phase_qcd}. 
Consequently,
observables which are sensitive to criticality related with a
spontaneous breaking of a chiral symmetry, in this fluctuations of net
baryon number and electric charge, should exhibit characteristic
properties governed by the universal part of the free energy density
\cite{karsch11:_probin_freez_out_condit_in,ejiri06:_hadron_fluct_at_qcd_phase_trans}. The
magnetic equation of states and cumulants of net charges at physical
quark masses have  been studied in first principle lattice QCD calculations
\cite{allton05:_therm_of_two_flavor_qcd,bazavov12:_fluct_and_correl_of_net,borsanyi12:_fluct_of_conser_charg_at,cheng09:_baryon_number_stran_and_elect},
as well as in effective chiral models
\cite{fukushima04:_chiral_polyak,sasaki07:_quark_number_fluct_in_chiral,sasaki07:_suscep_polyakov,stokic09:_kurtos_and_compr_near_chiral_trans,skokov10:_meson_fluct_and_therm_of,skokov10:_vacuum_fluct_and_therm_of_chiral_model,skokov11:_quark_number_fluct_in_polyak,friman11:_fluct_as_probe_of_qcd,asakawa09:_third_momen_of_conser_charg,herbst11:_phase_struc_of_polyak_quark,schaefer12:_qcd_critic_region_and_higher,wagner10:_effic_comput_of_high_order}.
Their properties which are obtained beyond mean-field level, have been shown to be  consistent with general
expectations originating from the $O(4)$ scaling.

The above  results have  opened new opportunity to verify the QCD
phase boundary  experimentally  by measuring fluctuations of conserved charges
\cite{karsch11:_probin_freez_out_condit_in,braun-munzinger11:_net_proton_probab_distr_in,kaczmarek11:_phase_qcd,braun-munzinger12:_net_charg_probab_distr_in,bazavov12:_fluct_and_correl_of_net,friman11:_fluct_as_probe_of_qcd,bazavov12:_freez_qcd},
and their probability distributions
\cite{morita12:_baryon_number_probab_distr_near,morita13:_net,nakamura13:_probin_qcd_phase_struc_by,morita14:_critic_net_baryon_number_probab}. This
is particularly the case, since the chiral pseudocritical line appears
near the phenomenological freezeout
line~\cite{karsch11:_probin_freez_out_condit_in}. Consequently, the
hadron resonance gas partition function constitutes  the regular part of a free
energy density of QCD in a hadronic phase, thus also, a reference for the non-critical
behavior of net charge fluctuations and their probability distribution
at the phase boundary
\cite{karsch11:_probin_freez_out_condit_in,braun-munzinger11:_net_proton_probab_distr_in,morita13:_net,morita14:_critic_net_baryon_number_probab}.

Cumulants of net baryon number fluctuations, quantified by the net protons, have recently been
explored in heavy ion collisions by  STAR Collaboration
\cite{aggarwal10:_higher_momen_of_net_proton,STAR_pn_2013}. The data
show deviations from the HRG reference, which are qualitatively
consistent with theoretical expectations based on the  $O(4)$ chiral
critical dynamics \cite{morita14:_critic_net_baryon_number_probab}. 
However, the role of different approximations and uncertainties
associated with the event-by-event measurements of fluctuations
remains to be clarified~\cite{kitazawa12:_reveal_baryon_number_fluct_from,kitazawa04:_relat_between_baryon_number_fluct,bzdak12:_accep_correc_to_net_baryon,bzdak13:_baryon_number_conser_and_cumul,nahrgang14:_impac_of_reson_regen_and}.

In particular, STAR measurements of the Beam Energy Scan program at the
Relativistic Heavy Ion Collider,  were carried out at
midrapidity  and within the transverse momentum range $0.4 < p_T < 0.8$ GeV/c.

The criticality related with the chiral symmetry restoration is
dominantly governed by soft momentum modes. One expects, that
produced particles with low momenta, carry information on such soft
mometum modes in an interacting medium. Although, there is no direct
one-to-one connection between the momentum scale in the interacting
medium and the momentum of emitted particles, nevertheless, one expects that imposing
cuts on the latter also restricts access to the soft modes. 
Consequently,  cuts imposed on particle momenta can implicitly influence properties of
cumulants of conserved charges near phase boundary. 

The main objectives of this paper is to study how the momentum cuts  can
modify critical properties of different cumulants of the net baryon
number  in the $O(4)$ universality class. Our studies are carried out
within the quark-meson (QM) model. In order to correctly account for
the $O(4)$ scaling properties of different physical observables near the
chiral transition we apply the functional renormalization group (FRG)
method
\cite{wetterich93:_exact_flow_equat,berges02:_non_pertur_renor_flow_in,delamotte07}.

We use the momentum scale dependence of the FRG flow equation
\cite{stokic10:_frg_scaling,nakano10:_fluct_and_isent_near_chiral_critic_endpoin,schaefer07:_suscep_near_qcd_tri_critic_point,Kamikado}
to examine how the suppression of the momentum modes in the infrared
and ultraviolet regimes modifies generic properties of the net-baryon
number fluctuations in the $O(4)$ universality class.

We show, that the pion mass $m_\pi$ is a natural  soft momentum  scale
at which cumulants are saturated at their critical values, whereas
for scales larger than $2m_\pi$ the characteristic $O(4)$ structure of
the higher order cumulants get lost. We also show, that the
restriction of momentum modes in the ultraviolet regime can as well
deflect the $O(4)$ structure of the net baryon number fluctuations.

Our  results indicate, that when measuring fluctuations of the net
baryon number in heavy ion collisions to search for  partial restoration
of the  chiral symmetry or critical point, a  detailed  dependence of the
results on kinematical cuts has to be examined.

This paper is organized as follows: in the next section, we introduce
the quark-meson model and its critical  properties.
In Sec.~\ref{sec:cm}, we present results on momentum scale dependence of
  different cumulants and their ratios. Section \ref{sec:summary} is
devoted to the concluding remarks.

\section{Quark-Meson model and fluctuations}

We employ the two-flavor quark-meson model to explore the momentum scale
dependence of the net quark number flucutations at finite temperature
and density. The quark-meson model is an effective realization of low energy
properties of QCD in which chiral symmetry breaking is described by
$O(4)$ meson multiplet $\phi = (\sigma, \vec{\pi})$ coupled to quark
fields $q$ with Yukawa coupling constant $g$. The QM model Lagrangian reads
\begin{align}
 \mathcal{L} &= \bar{q} [ i\gamma_\mu \partial^\mu - g (\sigma +
 i\gamma_5 \vec{\tau}\cdot \vec{\pi})]q + \frac{1}{2}(\partial_\mu \sigma)^2
+ \frac{1}{2}(\partial_\mu \vec{\pi})^2 - U(\sigma,\vec{\pi}),
\end{align}
where $U(\sigma, \vec{\pi})$ denotes the mesonic potential
\begin{equation}
 U(\sigma,\vec{\pi}) = \frac{1}{2}m^2 \phi^2 + \frac{\lambda}{4}\phi^4-h\sigma.
\end{equation}
The $O(4)$ chiral symmetry  is spontaneously broken to $O(3)$ yielding
$\langle \sigma \rangle \neq 0$
when $m^2 < 0$.  The explicit symmetry breaking term  $-h\sigma$ with
$h=f_\pi m_\pi^2$ gives the nonzero pion mass.

\subsection{Flow equation for quark-meson model at finite temperature
  and density}

The functional renormalization group (FRG) approach provides an efficient
method to evaluate the effective potential, which accounts for quantum
fluctuations \cite{wetterich93:_exact_flow_equat,berges02:_non_pertur_renor_flow_in,delamotte07,gies:_intro_rg}.

We introduce a scale dependent effective action $\Gamma_k[\phi,q]$, which
becomes the classical action $S$ at the ultraviolet cutoff scale
$\Lambda$,  and the full quantum effective action $\Gamma[\phi,q]$ in the
$k\rightarrow 0$ limit
\begin{align}
 \Gamma_\Lambda = S, ~~
 \Gamma_{k\rightarrow 0}= \Gamma. \label{eq:effectiveaction_lambda}
\end{align}
The evolution of  $\Gamma_k[\phi,q]$ with the renormalization scale $k$
is given by the following flow equation
\cite{wetterich93:_exact_flow_equat}
\begin{align}
 \partial_k \Gamma_k[\phi,q]
= - \text{Tr}\left[ \partial_k R_{kF}(p)(R_{kF}(p) +
		\Gamma_k^{(2,0)})^{-1} \right] 
 +\frac{1}{2}\text{Tr}
  \left[ \partial_k R_{kB}(p) (R_{kB}(p)+\Gamma_k^{(0,2)})^{-1}
 \right],
 \label{eq:frg_floweq}
\end{align}
where $\partial_k \equiv \partial/\partial k$, and the trace runs over
the internal momentum $p$,  as well as  spinor, color and flavor indices.
The $R_{k,B}(p)$ is an arbitrary cutoff function
which suppresses propagations of the bosonic modes with $p < k$,
originating from inserting a mass-like term,
$\frac{1}{2}\int \frac{d^D p}{(2\pi)^D} R_{k,B}(p)\phi(p)\phi(-p)$, into the
action. The fermionic counterpart, $R_{k,F}(p)$ is  introduced in
a similar fashion.
The $\Gamma_{k}^{(a,b)}$ in Eq. ~\eqref{eq:frg_floweq},  denotes the
$a-$times fermionic and $b-$times bosonic functional derivatices of
$\Gamma_k[\phi,q]$.

Owing to the scale dependent two-point functions,
the flow equation \eqref{eq:frg_floweq} has the one-loop structure with
an insertion of $\partial_k R_{k,B(F)}(p)$ which has a strong peak at
$p= k$. At finite temperature, following the standard imaginary time
formalism, the integral over the loop momentum $p$ is replaced by a
Matsubara sum.

To solve the flow equation, we employ the following optimized regulator functions
\cite{litim01:_optim_renor_group_flows}
\begin{align}
 R_{k,B}^{\text{opt}}(\boldsymbol{p}) &=
 (k^2-\boldsymbol{p}^2)\theta(k^2-\boldsymbol{p}^2)\label{eq:opt_regulator}\\
 R_{k,F}^{\text{opt}}(\boldsymbol{p}) &= (p+i\mu \gamma^0)\left(
 \sqrt{\frac{(p_0+i\mu)^2 + k^2}{(p_0+i\mu)^2 + \boldsymbol{p}}}-1\right)\theta(k^2-\boldsymbol{p}^2)\label{eq:opt_reg_fermion}
\end{align}
In the integration over the internal momentum $p$, the cutoff
functions $R_{k,B(F)}(p)$ in a full propagator plays a role of a mass
below $p < k$, and its derivative $\partial_k R_k(p)$ implements an
integration of momentum shell, as in an original Wilsonian idea.
By integrating the flow equation Eq.~\eqref{eq:frg_floweq}, from
$k=\Lambda$ to $k\simeq 0$, one gets the full effective action,  which includes quantum fluctuations.

The use of spatial momenta in the regulator functions in
Eqs.~\eqref{eq:opt_regulator} and \eqref{eq:opt_reg_fermion} allows 
summation over all Matsubara modes.
Thus we deal with the 3d system in thermal equilibrium
entirely during the scale evolution. 
In principle, one can employ an Euclidean-invariant form of the
regulators by replacing $\boldsymbol{p}^2$ with $p^2$. However, such a
procedure naturally includes the cutoff also in the Matsubara modes
which induces various difficulties and the dimensional reduction can be
achived only in $T/k \gg 1$
\cite{litim01:_optim_renor_group_flows,Schaefer99}.
Since our main objective is to investigate the influences of the three momentum cut on
$O(4)$ behavior of fluctuations of the conserved charges,
 we have not introduced the cut in the Matsubara modes.
In this way we avoid the modifications of  the thermal medium properties, in particular,
its Boltzmann momentum distribution.

The flow equation \eqref{eq:frg_floweq} for the scale dependent
effective action, includes the two-point function. Formally one can
obtain the flow equation for the scale-dependent two-point function by
taking the functional derivatives with respect to the fields. As  the
FRG flow includes higher order correlation functions, the flow equation
exhibits an infinite hierarchy, which one needs to truncate to solve it.

The evolution of the  $k$-scale dependent quantities, the so-called RG
trajectory, depends on the choice of the  regulator $R_k$ and how to
truncate the hierarchy,  by construction. 

To investigate critical phenomena, which are governed by soft modes,
it is convenient to assume,  that the field $\phi(x)$ varies slowly.
Then,  one can put an ansatz for the scale dependent effective action,
based on the derivative expansion. To leading order,  and ignoring field
renormalization,  the ansatz reads
\begin{equation}
 \Gamma_k[\phi,q] = \int d^Dx \, \left[ \bar{q} [ i\gamma_\mu \partial^\mu - g (\sigma +
 i\gamma_5 \vec{\tau}\cdot \vec{\pi})]q + U_k(\phi(x)) +
				\frac{1}{2}(\partial_\mu \phi(x))^2 \right],
\end{equation}
which is called, the  local potential approximation.
This approximation together with the optimized cutoff function has been
shown to produce the $O(4)$ criticality in the QM
model~\cite{stokic10:_frg_scaling}. Thus we expect that different
regulators and truncation schemes result only in small quantitative
difference.

Putting this ansatz and the regulator functions \eqref{eq:opt_regulator}
to the flow equation \eqref{eq:frg_floweq}, one obtains a
differential equation for the effective potential $U_k$ in a closed
form. Introducing a field variable $\rho \equiv \phi^2/2 = (\sigma^2 +
\vec{\pi}^2)/2$,
the flow equation for the scale dependent thermodynamic potential density
$\Omega_k = T\Gamma_k/V$ is given by
\begin{equation}
 \begin{split}
  \partial_k \Omega_k(\rho;T,\mu)=\frac{k^4}{12\pi^2}\left[
  \frac{3}{E_\pi}\left\{1+2n_B(E_\pi)\right\}
  +\frac{1}{E_\sigma}\left\{1+2n_B(E_\sigma)\right\} \right. \\
  \left. -\frac{2\nu_q}{E_q}\left\{
  1-n_F(E_q+\mu)-n_F(E_q-\mu)\right\}
  \right]\label{eq:floweq}.
  \end{split}
\end{equation}
The first, second, and third terms stand for $\pi$, $\sigma$, and quark
contributions, as seen from the corresponding thermal distribution
functions $n_B$ and $n_F$ with the quasiparticle energies
 $E_a = \sqrt{k^2 + m_{a,k}^2}$, where $a=\pi,\sigma$ or $q$.
$\nu_q = 2N_f N_c$ denotes the quark degeneracy.

The scale dependent effective mass of mesons are related to
the effective potential with the explicit breaking term being removed,
$\bar{\Omega}_k = \Omega_k + h\sqrt{\rho_k}$,
\begin{align}
 m_{\pi,k}^2&= \Omega_k^\prime \\
 m_{\sigma,k}^2 &= \bar{\Omega}_k^\prime + 2\rho_k \bar{\Omega}_k^{\prime\prime},
\end{align}
where $\bar{\Omega}_k^\prime = \partial \bar{\Omega}_k/\partial \rho$.
The dynamical quark mass is directly related to the order parameter
\begin{equation}
 m_{q,k} = g \sigma_k,
\end{equation}
where the Yukawa coupling  $g=3.2$ is fixed  to get
$m_{q,0}=300$MeV at $T=0$.

The flow equation \eqref{eq:floweq} is solved numerically with a Taylor
expansion method \cite{stokic10:_frg_scaling} in which the
scale dependent potential is expanded around its minimum $\rho_k$,  up to the third
order in $\rho$,  as
\begin{equation}
 \bar{\Omega}_k(\rho) = \sum_{n=0}^{3}\frac{a_{n,k}}{n!} (\rho-\rho_{k})^n.
\end{equation}
The coefficients $a_{n.k}$,  and the minimum $\rho_{k}$,
follow the following flow equations
\begin{align}
 d_k a_{0,k}&= \frac{h}{\sqrt{2\rho_k}}d_k \rho_k + \partial_k
 \Omega_k, \nonumber\\
 d_k \rho_k&=
 -\frac{1}{c/(2\rho_k)^{3/2}+a_{2,k}}\partial_k\Omega_k^\prime, \nonumber\\
 d_k a_{2,k}& = a_{3,k} d_k \rho_k + \partial_k
 \Omega_k^{\prime\prime}, \nonumber\\
 d_k a_{3,k}&= \partial_k \Omega_k^{\prime\prime\prime},\label{eq:flow_expanded}
\end{align}
where  $d_k \equiv d/dk$,  and $a_{1,k}$ is eliminated by making use of the scale independent
relation $h\equiv a_{1,k}\sqrt{2\rho_k}$.
The initial condition at an ultraviolet cutoff scale $\Lambda=1.0$ GeV
is  chosen so as to satisfy the requirement
\eqref{eq:effectiveaction_lambda},  and to reproduce the vacuum physics.
Therefore, we set $a_{3,\Lambda}=0$,  whereas  $a_{2,\Lambda}$ and
$\rho_{\Lambda}$ are fixed such as  to reproduce
$\sigma_{k=0}(T=\mu=0)=f_\pi=93$MeV and
$m_\sigma=640$MeV with $m_\pi=135$MeV. While $a_{0,\Lambda}$  gives
a constant shift in thermodynamic potential density, the lack of
contributions from degrees of freedom above the ultraviolet cutoff
scale,  causes an unphysical behavior in thermodynamic quantities at
high temperature \cite{braun04:_linkin_qcd,
skokov10:_meson_fluct_and_therm_of}. Thus, we include such contribution
by integrating the flow equation from $k=\infty$ to $k=\Lambda$ for a
non-interacting massless quarks and gluons,
\begin{equation}
 \partial_k \Omega_k^{\Lambda}(T,\mu) = \frac{k^3}{12\pi^2} \{
  2(N_c^2-1)[1+2n_B(k)] 
 -\nu_q [1-n_F(k+\mu)-n_F(k-\mu)]\}.
\end{equation}
The pressure of the system  is  then obtained as,
$p(T,\mu) =-\Omega_{k\rightarrow 0}$.

\section{Momentum scale and  criticality}\label{sec:cm}

\subsection{Order parameter and meson masses}

In  the chiral limit, and at the moderate values of quark chemical
potential,  the QM model is well known to exhibit the second order phase
transition  in the $O(4)$ universality class which terminates at the
critical end point, and then follows as the first order transition. For
a  physical  pion mass  the  $O(4)$ phase transition is washed out
and becomes a smooth crossover. Nevertheless, due to smallness   of the
light quark  masses the crossover is  constituted as the  pseudocritical
line along which the physical observables follow the scaling properties
of the $O(4)$ universality class.

\begin{figure}[ht]
 \centering
  \includegraphics[width=0.6\textwidth]{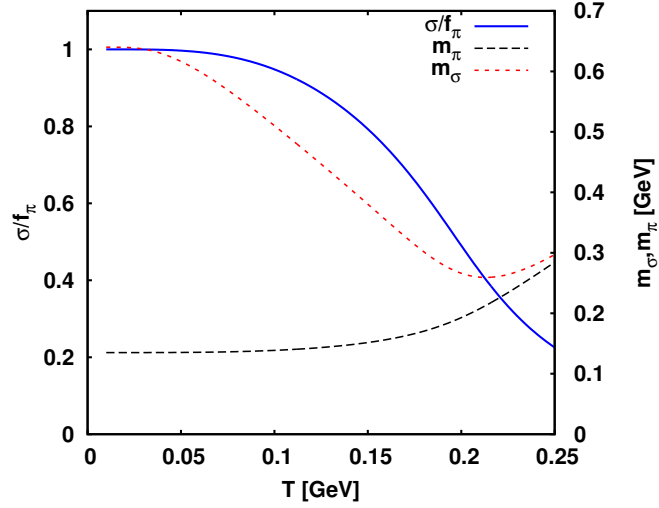}
 \caption{The order parameter, the sigma and pion mass as a function
 of temperature,  calculated in the quark-meson model within the
 functional renormalization group approach.}
 \label{fig:sigma-T}
\end{figure}


 \begin{figure*}[ht]
  \centering
  \includegraphics[width=0.49\textwidth]{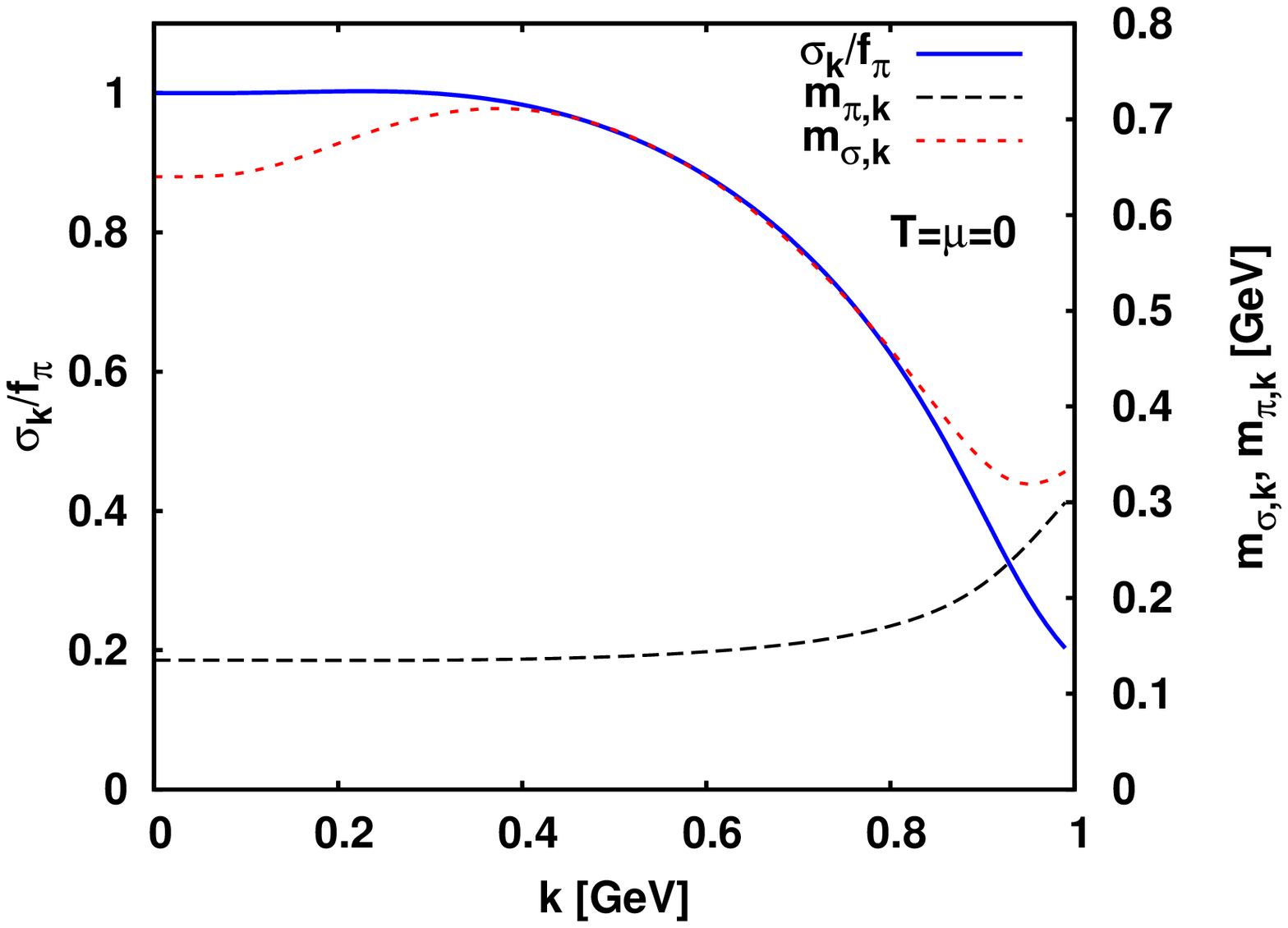}
  \includegraphics[width=0.49\textwidth]{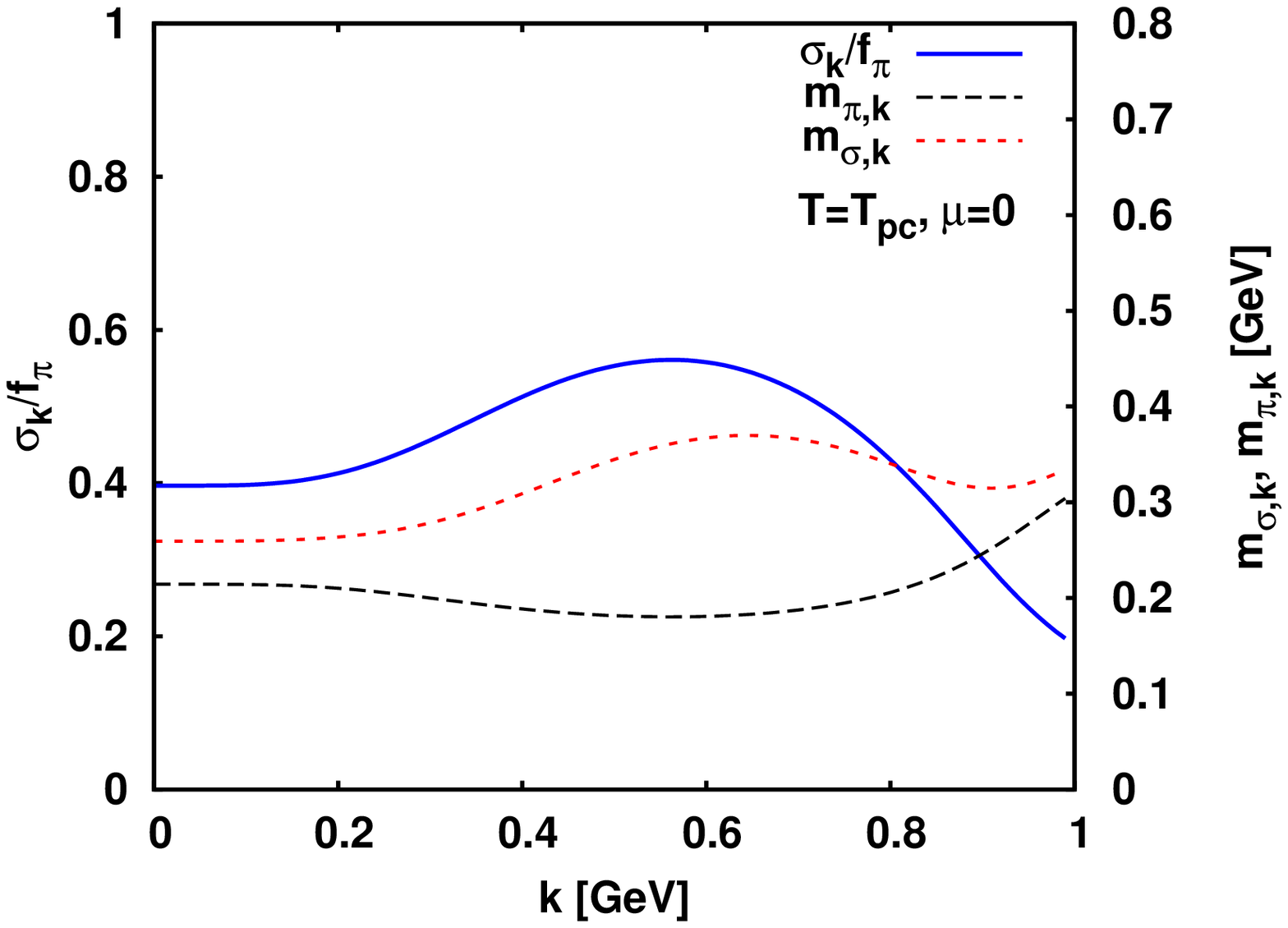}
  \caption{The momentum scale dependence of the order parameter, the
  sigma and pion mass in the quark-meson model obtained from the
  renormalization group flow equation. The left-hand figure
  corresponds to $T=0$  and the right-hand figure to $T=T_{pc}$, where
  $T_{pc}$ is the pseudocritical temperature.}
  \label{fig:sigma_kdep}
 \end{figure*}

In the QM model,  the order parameter of the chiral phase transition,
is the expectation value of the $\sigma$-field.
In Fig.~\ref{fig:sigma-T}, we show melting of   the order parameter with
temperature obtained by solving the flow equation
\eqref{eq:flow_expanded}. The chiral symmetry  is spontaneously broken
in a vaccum and is partially restored  in a medium at  some
pseudocritical temperature  $T=T_{pc}$. The values  of  $T_{pc}$ for
different quark chemical potential,  can be identified as a peak
position of the susceptibility  of the order parameter or as the minimum
of the sigma mass.

Figure \ref{fig:sigma-T} also shows
 the temperature dependence of the
sigma and pion mass. The sigma mass is decreasing with temperature,
whereas the pion mass is increasing towards  $T_{pc}$,  where it is
approximately degenerate with the sigma mass. In the chiral limit the
chiral condensate vanishes at the critical point  where the sigma and
the pion masses coincide.

The results shown in Fig.~\ref{fig:sigma-T} were obtained within FRG
approach, where all momentum modes up to $k=0$ were integrated out,
thus  the thermal and quantum fluctuations have been included within the
local potential approximation.
However, the results of the RG-flow equations for the evolution of
different physical observables with the infrared cutoff  $k\neq 0$
can be also used to study the chiral symmetry breaking.

In Fig.~\ref{fig:sigma_kdep}-left, we show the momentum scale dependence of
the order parameter,  the  pion and the sigma mass at $T=0$.  At large
momentum  scale $k\sim \Lambda$, the chiral symmetry is approximately
valid, which is reflected in Fig.~\ref{fig:sigma_kdep} by the small value of the order
parameter and almost degenerate pion and sigma masses. When decreasing
the scale bellow the ultraviolet cutoff $\Lambda$, there is a rapid
grow of the order parameter towards its vacuum value. There is also a
corresponding increase of the sigma mass after reaching a  minimum  at
$k\sim k_{ch}$, where $k_{ch}\simeq 900$ MeV   constitutes the momentum
scale for an approximate chiral symmetry breaking.

The pion mass is seen in Fig.~\ref{fig:sigma_kdep}-left to decrease
monotonically with decreasing momentum scale.
On the other hand, the sigma mass is a non-monotonic function of  $k$,
as it reaches a maximum at $k\simeq 400$ MeV and then decreases towards the
vacuum value. This property of sigma mass is consistent with previous
finding in Ref.~\cite{braun05:_volum}, and can be attributed to differences
between  the constituent quark and pion masses. A decrease of  $m_\sigma$
bellow  $k<400$ MeV is due to presence of the light pion.

At finite temperature, the  scale dependence of the order parameter,
the sigma and pion masses is qualitatively similar to  the $T=0$ case. In
Fig.~\ref{fig:sigma_kdep}-right, we show the scale dependence of these
observables at $T=T_{pc}$, where the chiral symmetry is partially
restored. At large momenta $k\sim \Lambda$ the temperature effect is
negligible. At lower scales, however, the thermal and meson fluctuations
prevent the order parameter from growing to its vacuum value.  The sigma mass
and the order parameter  reaches their maximum   at $k\sim 600$ MeV, and
then  decrease towards   values at the pseudocritical temperature. The
pion exhibits a broad minimum at similar scale  $k\sim 600$ MeV and then
slightly increases towards $k=0$.

Thus, it is clear that the  RG-flow equations  for the evolution with
the infrared cutoff give a picture how  the chiral criticality developed
in a medium with the momentum  scale. Clearly, to reproduce the
expected $O(4)$ scaling of physical observables at the chiral crossover
one needs the scale evolution towards $k = 0$. Any restriction on
momentum scale in a medium modifies thermodynamically relevant
information on universal properties.

\begin{figure*}[ht]
 \centering
 \includegraphics[width=0.49\textwidth]{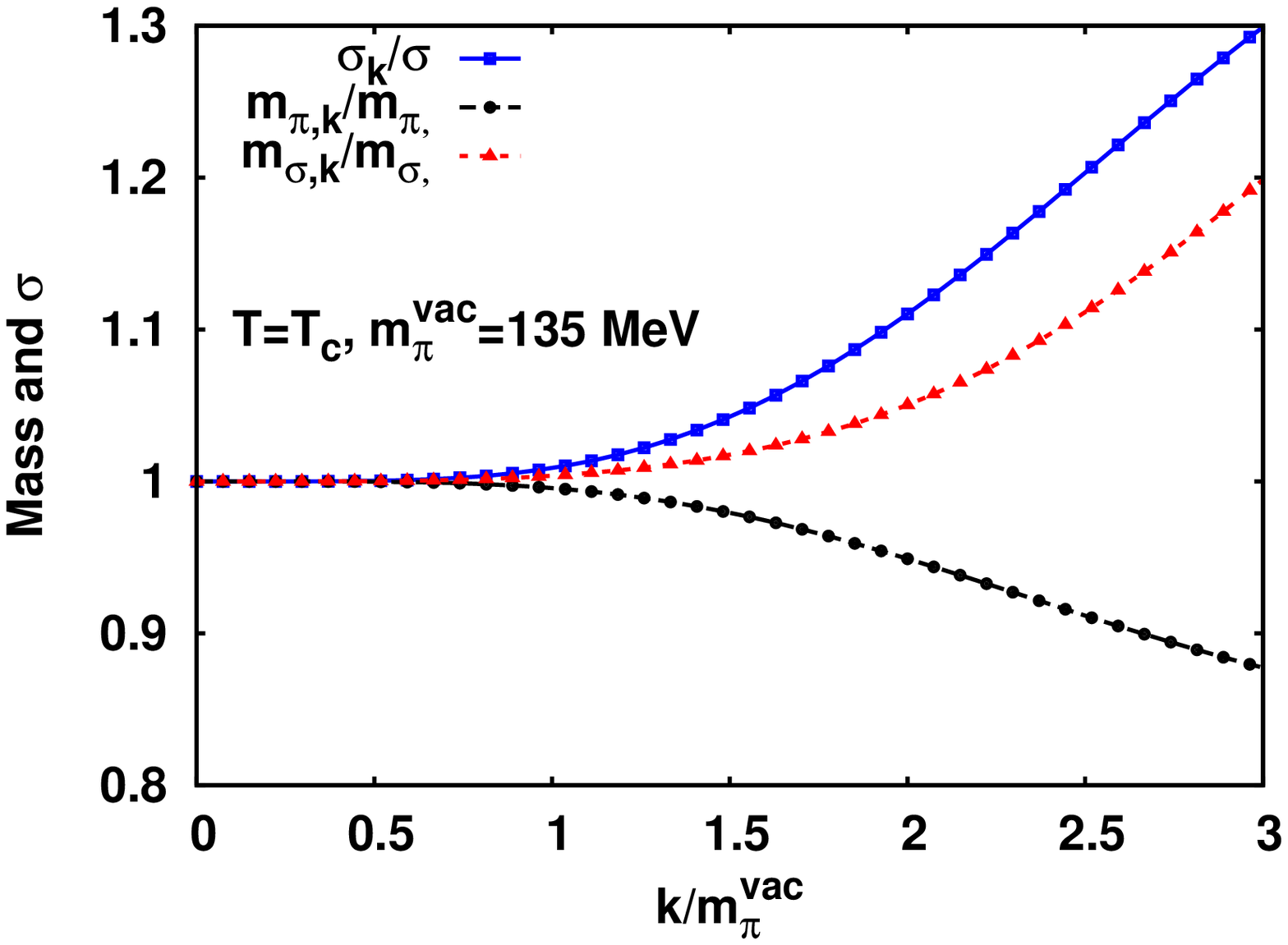}
 \includegraphics[width=0.49\textwidth]{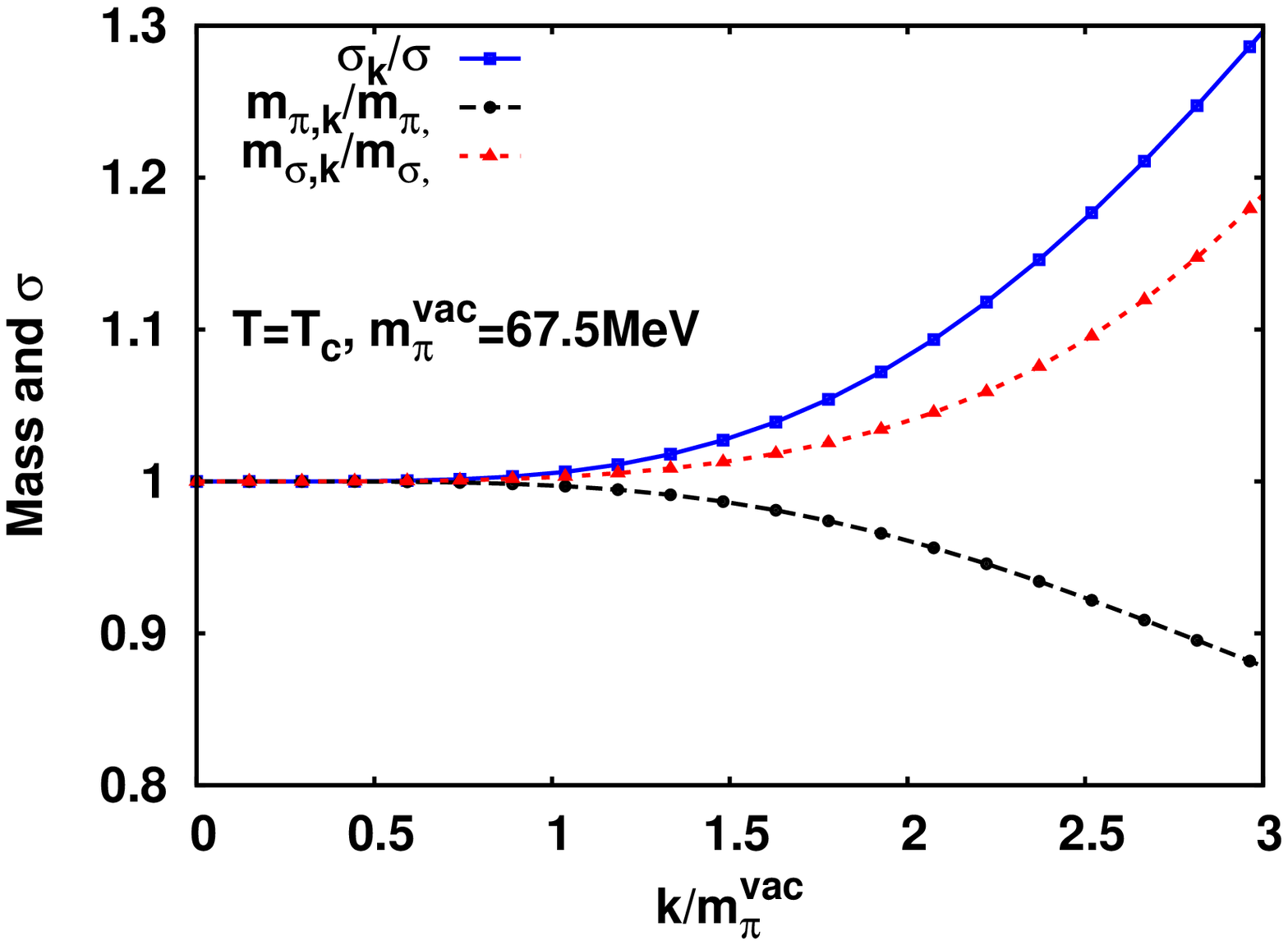}
 \caption{As in Fig.~\ref{fig:sigma_kdep}-right, but the quantities are
 normalized to their values at $k=0$ and the momentum scale is also
 normalized by the vacuum pion mass. The left-hand figure is
 calculated for the vacuum pion mass $m^{\rm vac}_\pi=135$ MeV, and
 the right-hand figure for $m^{\rm vac}_\pi=67.5$ MeV.  }
 \label{fig:mass-kovermpi}
\end{figure*}

To describe consequences of momentum scale cuts on the critical
properties of relevant observables  at the chiral crossover, we introduce
the scale dependent  ratios of such quantities calculated at the scale
$k$ and at $k=0$. Figure \ref{fig:mass-kovermpi} shows  ratios for the order parameter, the pion
and sigma masses at vanishing quark chemical potential and  for two
different vacuum pion masses. It is very transparent from
Fig.~\ref{fig:mass-kovermpi},  that at $T_{pc}$ the natural momentum
scale where these observables saturate at their critical values  is the
pion mass. This is seen by comparing the $m_\pi=135$ and $m_\pi=67.5$ MeV
cases.  One expects, that such a scale should be governed by the softest
mode in a system.~\footnote{In actual studies this is a quark mass
$m_q$. However, this is not an observable and the critical behavior,
\textit{i.e.,} divergent fluctuations of conserved charges, is due to the mesonic sector
since the light quark mass effect is already reflected in the mean field
approximation.}

Thus, if the momentum scale reaches $k\simeq m_\pi$, all these observables
decouple from the RG flow. One can also  conclude, that introducing any
momentum cut in a system at $k\leq m_\pi$ will not modify relevant $O(4)$
properties near chiral crossover. 


\subsection{The net quark number fluctuations and momentum cuts}

The sensitivity to the $O(4)$ criticality increases if the higher order
fluctuations of the order parameter or conserved charges are considered
at the chiral crossover.  Of particular phenomenological interests are
$n$th-order cumulants of the net quark  number $\chi_n$,  which have been
successfully quantified through the measurement of net proton number in
heavy ion collisions by STAR
Collaboration~\cite{aggarwal10:_higher_momen_of_net_proton,STAR_pn_2013}.
Theoretically, the $\chi_n$ are obtained as derivatives of
thermodynamic pressure with respect to the quark chemical potential,
\begin{equation}
 \chi_n(T,\mu) \equiv \frac{\partial^n [p(T,\mu)/T^4]}{\partial (\mu/T)^n}. \label{cumulants}
\end{equation}
In the vicinity of the chiral crossover, and due to remnants of the $O(4)$
criticality, the   $\chi_n$ receive contributions from the regular  and
the singular  part of the free energy density. Consequently, near
$T_{pc}$ one can decompose $\chi_n$=$\chi_n^s$+$\chi_n^r$,
correspondingly. Owing  to the $O(4)$ scaling, the  $\chi_n^s$  show a
strong dependence on the explicit symmetry breaking term $h$, the quark
mass
\begin{equation}
\chi_n^s \sim
\begin{cases}
- h^{(2-\alpha -n/2)/\beta\delta} f_f^{(n/2)}(T,\mu)
& \ {\rm for}\ \mu = 0\
 \\
-  \left( \frac{\mu}{T} \right)^n
h^{(2-\alpha -n)/\beta\delta} f_f^{(n)}(T,\mu)
&\ {\rm for}\ \mu > 0\, ,
\end{cases}
\label{scaling}
\end{equation}
where due to quark-antiquark symmetry at $\mu=0$, the first equation
holds only for even $n$. The  $\alpha,\beta$ and $\delta$ are critical
exponents in the $O(4)$ universality class, and $f^{(n)}$ is the $n$-th
order derivative of the $O(4)$ universal scaling function with respect
to the scaling variable
\cite{engels12:_scalin_funct_of_free_energ,friman11:_fluct_as_probe_of_qcd}.

As $\alpha=-0.2131 (34)$,  is negative in the  $O(4)$ universality class
\cite{engels12:_scalin_funct_of_free_energ},
the second and the fourth order cumulants of the net quark number fluctuations
are finite in the chiral limit at the chiral transition
temperature. From Eq. ~\eqref{scaling}, it is clear, that at $\mu=0$
the first divergent moment is obtained for $n= 6$, whereas at
$\mu > 0$   for $n= 3$.

At fixed $h$, the $T$- and $\mu$-dependence of $\chi_n^s$ is entirely
governed by derivatives of the $O(4)$  universal scaling function. The
characteristic feature of the sixth order cumulants at $\mu=0$ is that
at physical pion mass it can be negative near $T_{pc}$
\cite{friman11:_fluct_as_probe_of_qcd}. This makes $\chi_6$ as an ideal
observable of partial restoration of chiral symmetry in heavy ion
collisions at RHIC and LHC  if the chemical freezout appears near the
chiral crossover
\cite{karsch11:_probin_freez_out_condit_in,friman11:_fluct_as_probe_of_qcd}.


\begin{figure}[!ht]
 \centering
 \includegraphics[width=0.6\textwidth]{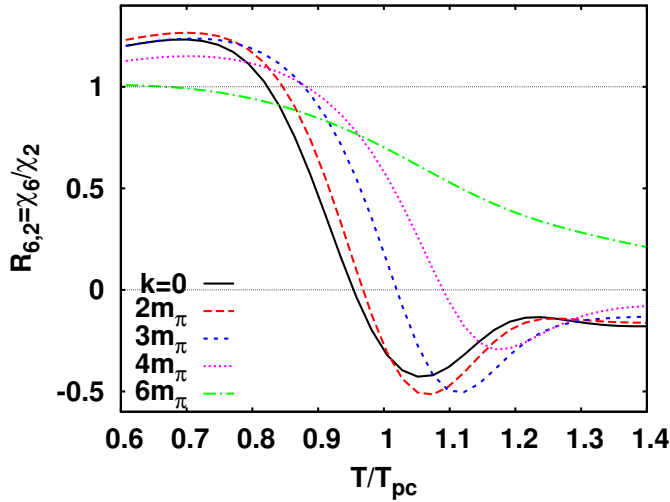}
 \caption{The momentum scale dependence of the sixth to second order
 cumulant ratio in the quark-meson model calculated within the
 functional renormalization  group method at finite temperature near
 the pseudocritical temperature $T_{pc}$. The momentum scales are
 expressed in units of a physical pion mass $m^{\rm vac}_\pi=135$ MeV. }
  \label{fig:c6c2-kcut}
\end{figure}

Figure \ref{fig:c6c2-kcut}  shows the ratio  $R_{6,2}=\chi_6/\chi_2$
near $T_{pc}$ at $\mu=0$ calculated in the QM model. Since $\chi_2$ is
governed only by the regular part, the observed strong variation of
$R_{6,2}$ and its negative structure, is entirely  due to remnant of the
$O(4)$ criticality in the sixth order cumulant.
Thus, the expected generic structure of the $O(4)$ scaling function in
$\chi_6$  is well reproduced  in the QM model within FRG approach
\cite{friman11:_fluct_as_probe_of_qcd,morita13:_net}, if the RG flow
terminates at $k=0$. Introducing the infrared momentum  cutoff $k>0$ can
clearly deflect criticality near $T_{pc}$.

In Fig.~\ref{fig:c6c2-kcut}, we show the $R_{6,2}$ by applying
different infrared momentum cuts in units of the vacuum pion mass.
The scale dependent net quark number fluctuations were calculated
numerically as derivatives of the scale dependent thermodynamic potential
density (\ref{eq:floweq}).
With increasing momentum scale, the suppression of $R_{6,2}$  near $T_{pc}$
due to $O(4)$ criticality  is weakened. For sufficiently large momentum
scale the characteristic negative structure  of this fluctuation ratio
disappears,  indicating that the singular part contribution to $\chi_6$
is suppressed. For $k>5m_\pi$,  the  $R_{6,2}$ shows a smooth
change from unity in the chirally  broken phase to the value expected in
the ideal massless quark gas. This indicates,  that bellow $T_{pc}$ and
for large $k$,  the structure of $\chi_6$ is governed by the Skellam
distribution,  and that a smooth decrease of  $R_{6,2}$ with $T$ is due to
decreasing quark mass  with temperature  and  quantum statistics
effect. However, if the momentum scale is smaller than $2m_\pi$,  then
a generic $O(4)$ structure  of    $R_{6,2}$  is preserved near
$T_{pc}$. For $k>2m_\pi$
the  $R_{6,2}$ is not anymore negative in the chirally broken phase just
bellow $T_{pc}$.

\begin{figure*}[ht]
 \centering
  \includegraphics[width=0.49\columnwidth]{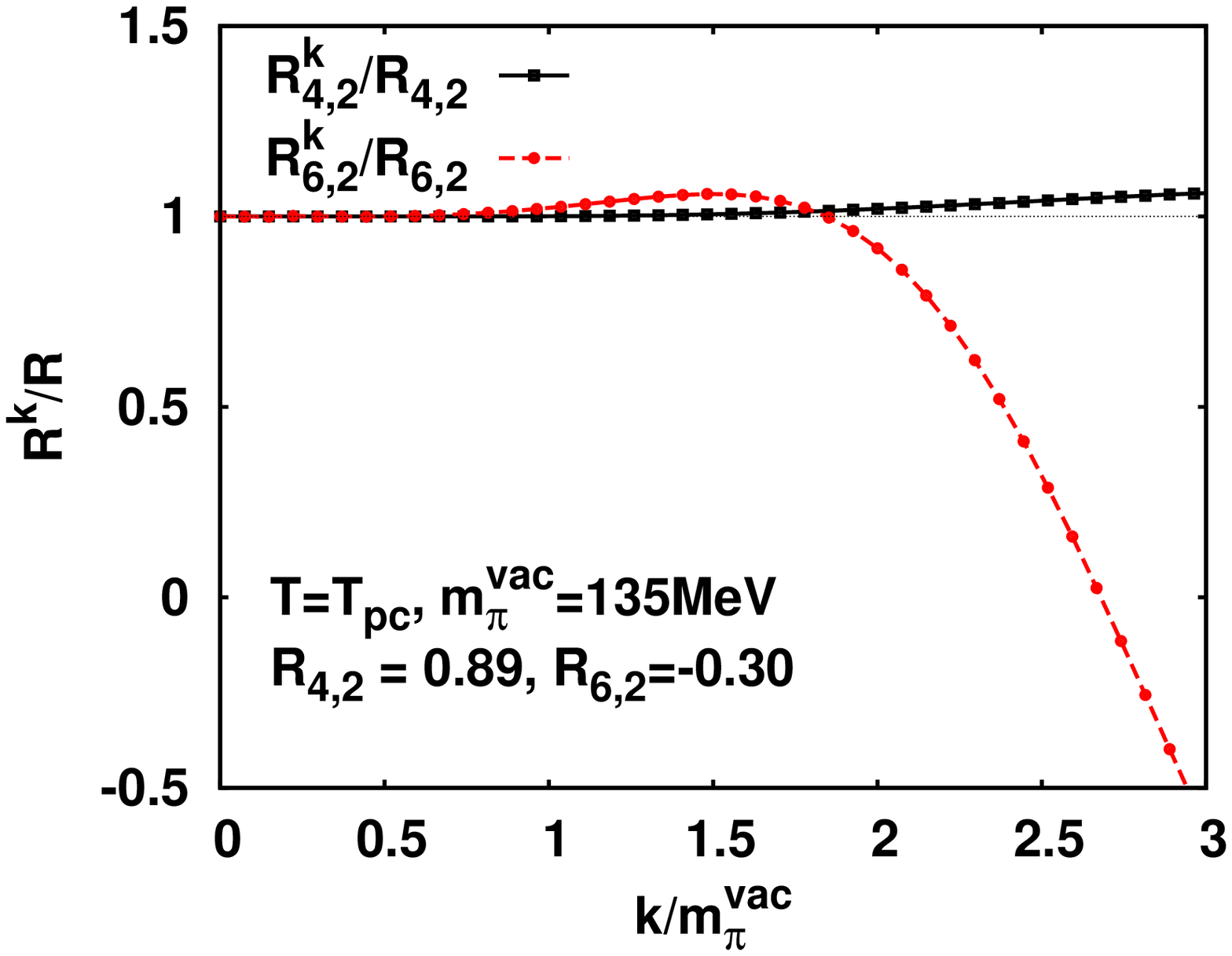}
  \includegraphics[width=0.49\columnwidth]{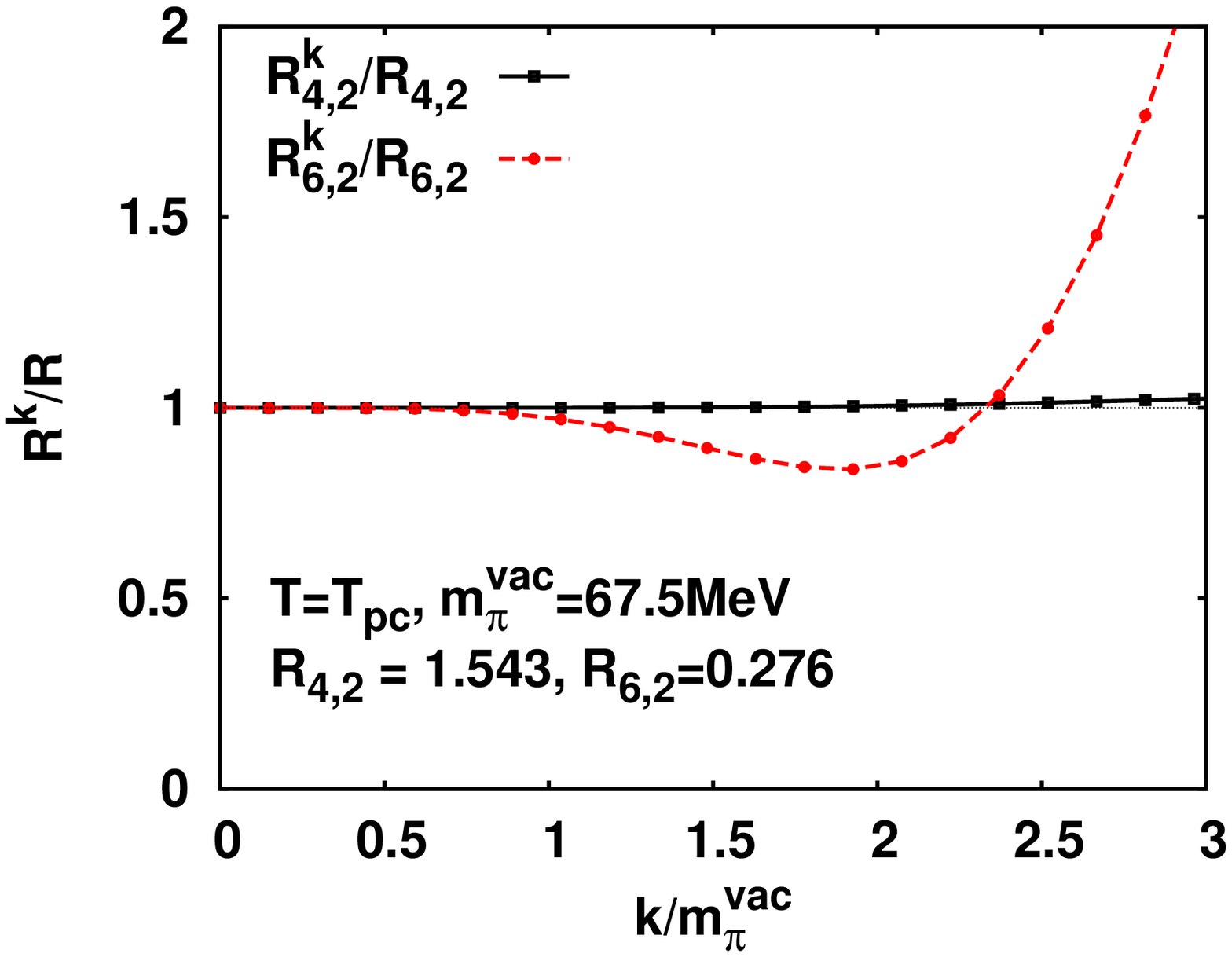}
  \caption{The net quark number cumulants ratio
  $R^k_{n,m}=\chi_n/\chi_m$ at the pseudocritical point $T_{pc}$,
  calculated within the FRG method at the soft momentum scale $k$, and
  normalized to their value at the scale $k=0$, indicated in the
  figures. The left-hand figure is calculated for the vacuum pion mass
  $m^{\rm vac}_\pi=135$ MeV, and  the right-hand figure for $m^{\rm
  vac}_\pi=67.5$ MeV.  }
  \label{fig:cum_Tpc_kdep}
 \end{figure*}

The change of $R_{6,2}$  with the infrared cutoff  at the chiral
crossover  temperature $T_{pc}$ is very transparent  when  considering
the ratio of $R_{6,2}$ calculated at momentum scale $k$ and that at
$k=0$. The corresponding results are shown in
Fig.~\ref{fig:cum_Tpc_kdep}  for different vacuum pion masses.  The pion
mass fixes the scale where $R_{6,2}$ saturates at its critical
value. Similarly to Fig.~\ref{fig:sigma_kdep}, if  the infrared
momentum scale reaches the softest mode which is approximately quantified
by the pion mass, then the sixth order cumulant  decouples form the RG
flow. Figure \ref{fig:cum_Tpc_kdep} also indicates, that for $k<2m_\pi$,
the  $R_{6,2}$ is weekly changing with infrared cutoff. Only for scales
larger than $2m_\pi$, the  $R_{6,2}$  at $T_{pc}$ is positive,  and the
characteristic negative fluctuation due to $O(4)$ criticality is
lost. At smaller pion mass $m_\pi\simeq 65$ MeV, the  $R_{6,2}$
calculated up to  $k=0$  is  positive at the $T_{pc}$ \footnote{The
smaller pion mass is, the sharper crossover transition becomes. Thus $R_{6,2}$
also exhibits a sharper change around $T_{pc}$. The temperature where
$R_{6,2}<0$,   shifts to higher values,  since in the chiral limit $R_{6,2} >0$ at
$T < T_{c}$,   and is positively or negatively divergent if
$T \rightarrow T_c$ from below  or above, respectively.} This implies,
that the normalized ratio shown in Fig. 5-right,  increases with $k$ for
$k>2m_\pi$.

In Fig.~\ref{fig:cum_Tpc_kdep},  the scale dependence of $R_{4,2}$ is
also shown. Since at vanishing chemical potential, both $\chi_4$ and
$\chi_2$ are non-critical at $T_{pc}$, $R_{4,2}$ is governed entirely by
the regular part of the free energy. Consequently, $R_{4,2}$ is almost
scale independent, as seen in this figure.

 \begin{figure}[ht]
  \centering
  \includegraphics[width=0.6\textwidth]{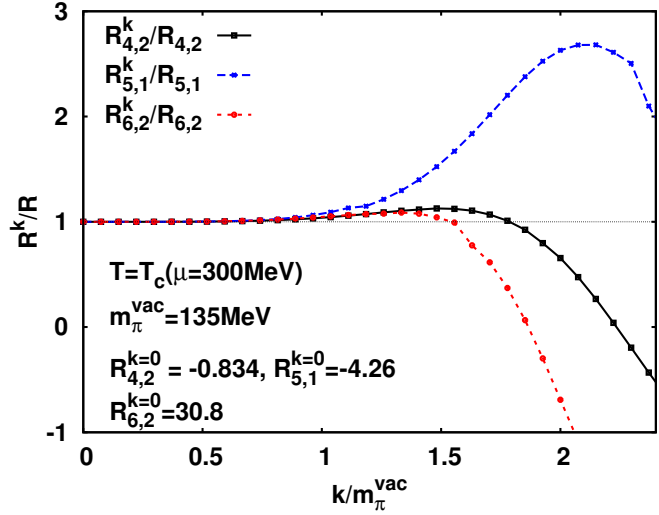}
  \caption{ As in Fig. 5-left,  but the calculations are done at the
  pseudocritical temperature $T=125$MeV for large finite quark chemical
  potential,  $\mu_{pc}=300$ MeV.}
  \label{fig:cum_tpc_nearcp}
 \end{figure}

At finite chemical potential, already the third and higher order
cumulants diverge at the chiral phase transition in the chiral
limit. Consequently, for finite pion mass all $\chi_n$ with $n \geq 3$
are influenced by the $O(4)$ criticality, thus should be also sensitive
to the momentum scale at which they are calculated.
In Fig.~\ref{fig:cum_tpc_nearcp}, we show ratios of
$R_{n,m}=\chi_n/\chi_m$,  calculated at momentum scale $k$ and at
$k=0$, for different orders of  cumulant. These normalized ratios are
evaluated at the chiral crossover point where the chemical potential
is $\mu_{pc}=300$ MeV and $T_{pc}=125$ MeV. Similarly, as at $\mu=0$,
all  $R_{n,m}$, shown in Fig.~\ref{fig:cum_tpc_nearcp}, saturate at
their pseudocritical values if the momentum scale $k$ reaches the pion
mass. For scales $m_\pi < k < 1.5m_\pi$, deviations of
$R_{3,1}, R_{4,2}$ and  $R_{6,2}$ from their critical values are small.
The $R_{5,1}$, however, exhibits much stronger scale
dependence. Comparing the flow of $R_{6,2}$ at finite and vanishing
$\mu$, one concludes that at $\mu\neq 0$, there is a stronger
sensitivity  of this cumulants ratio to the soft momentum scale.

So far,  effects of \textit{infrared} momentum cut on the fluctuation
observables have been discussed. Although,  the criticality associated
with partial restoration of chiral symmetry is governed by soft
momentum modes,  thus is related to the infrared cutoff, the ultraviolet
cutoff also influences the fluctuations of the conserved charges.

 \begin{figure*}[ht]
  \centering
  \includegraphics[width=0.49\textwidth]{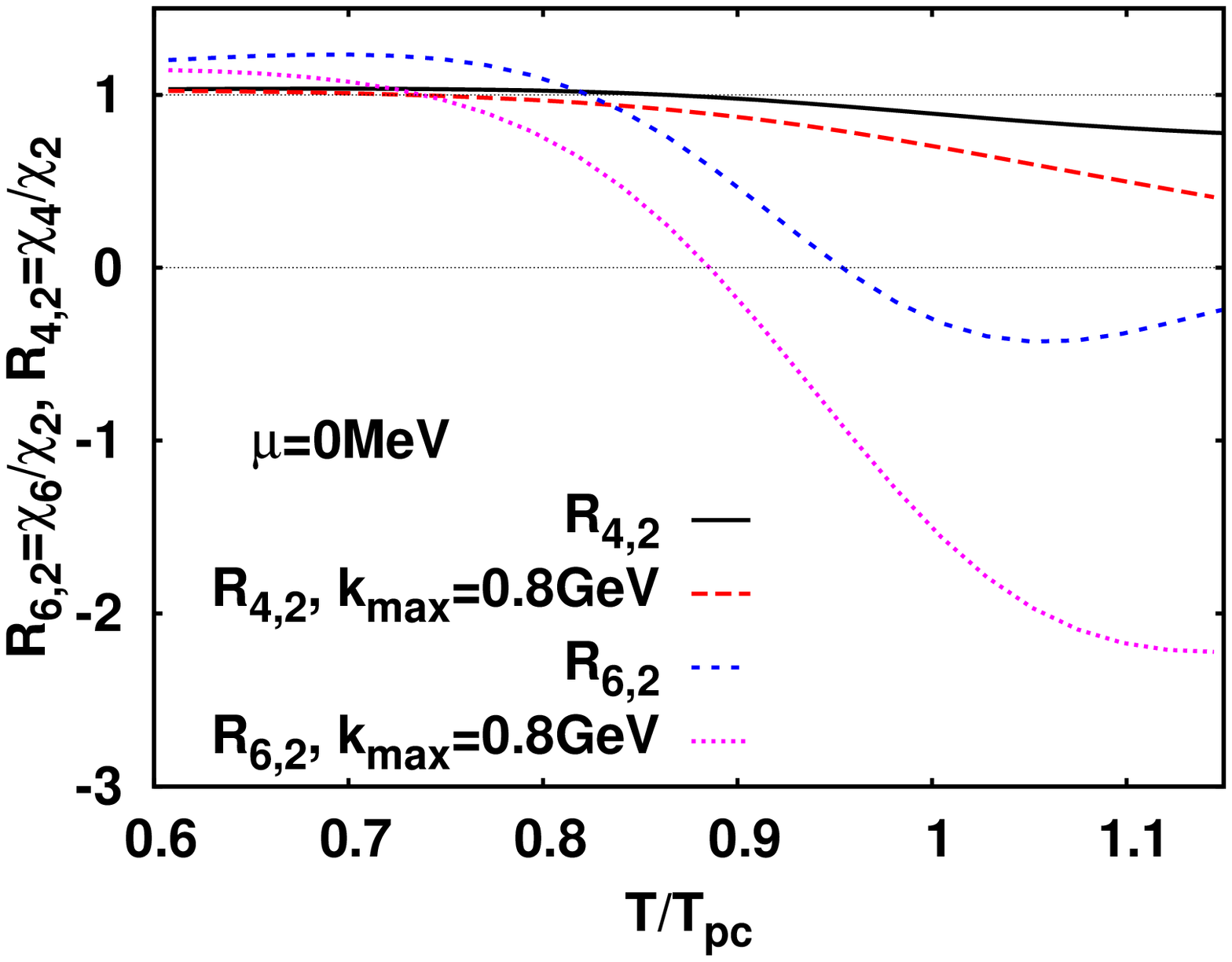}
  \includegraphics[width=0.49\textwidth]{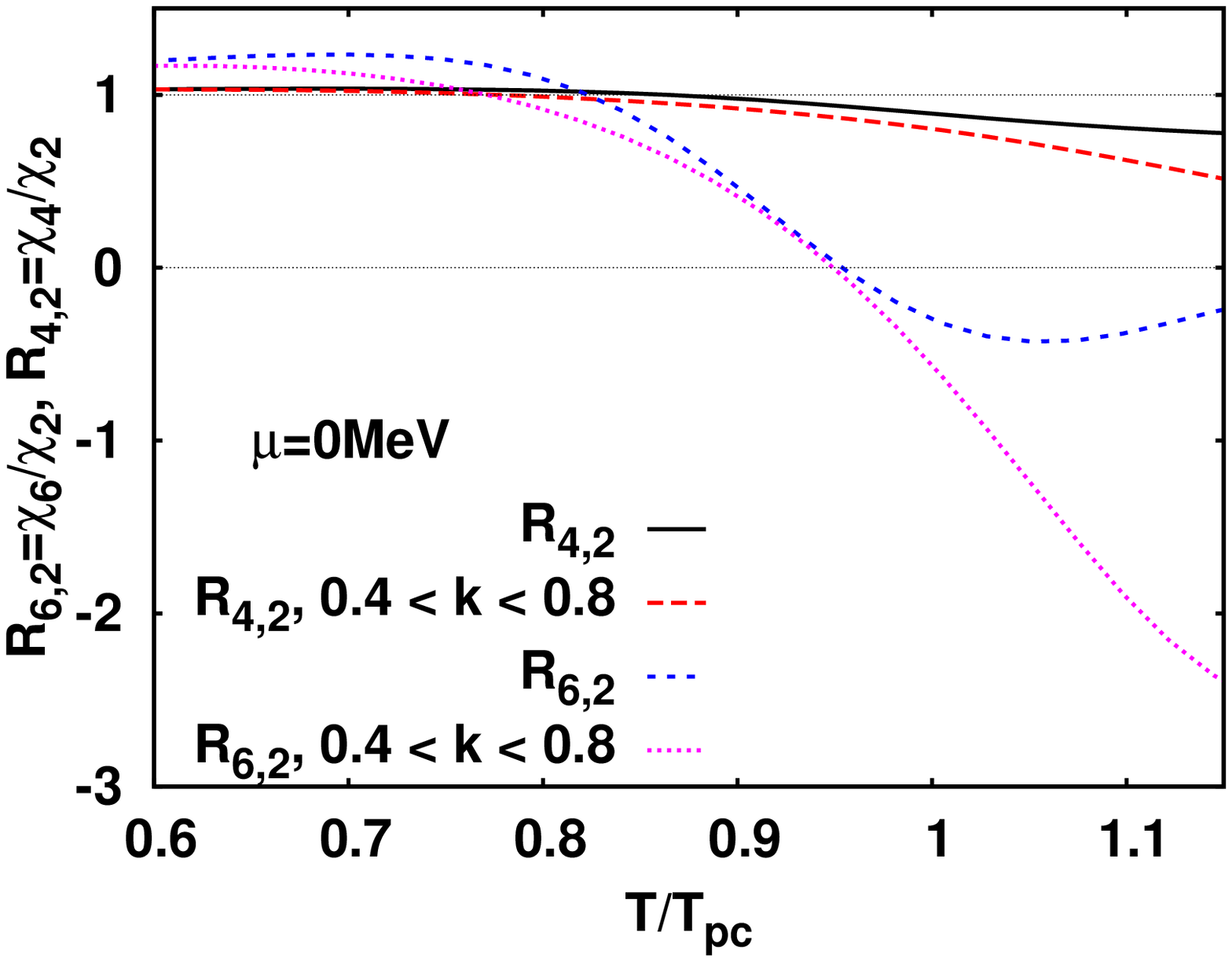}
  \caption{Temperature dependence of the net quark number cumulants
  ratio, $R_{n,m}=\chi_n/\chi_m$. In the left hand figure: the
  $R_{n,m}$,  calculated in the quark-meson model in the FRG flow within
  full momentum range is compared with the corresponding result with the
  ultraviolet mometum cut $k_{\rm max}=0.8$ GeV. In the right hand
  figure:  the full FRG result for  $R_{n,m}$ is compared with the
  corresponding result obtained in the momentum interval $0.4<k<0.8$
  GeV. }
  \label{fig:cum62_cutoff}
 \end{figure*}

Figure \ref{fig:cum62_cutoff}-left displays the temperature dependence
of the cumulants ratios $R_{4,2}$ and $R_{6,2}$ with and without the
ultraviolet cutoff $k_{\text{max}}=0.8$GeV. The calculation was done by
setting the initial momentum scale to $k_{\text{max}}$ without changing
the vacuum physics, such that flow of the observables follow the same
trajectory. The behavior of $R_{4,2}$ was already discussed in
Ref.~\cite{skokov10:_meson_fluct_and_therm_of}. 
The absence of high momentum contribution implies, that  $\chi_2$ and
$\chi_4$ turn to decrease above $T_{pc}$, leading to suppression at high
temperature. As seen in Fig.~\ref{fig:cum62_cutoff}, $R_{6.2}$ follows
the same trend.

Results employing,  both infrared and ultraviolet cutoffs, $0.4 < k < 0.8$
GeV,  are shown in Fig.~\ref{fig:cum62_cutoff}-right.
The effects of infrared  and  ultraviolet  cutoffs on $R_{4,2}$, applied
separately,  were shown in Figs.~\ref{fig:cum_Tpc_kdep}  and
~\ref{fig:cum62_cutoff}-left  to be small. Consequently, $R_{4,2}$
is seen in Fig.~\ref{fig:cum62_cutoff}-right, to be also weakly
sensitive  if the momentum scales are constrained in both limits,
simultaneously.  
However,  the  structure of $R_{6,2}$ is strongly changed.
Its characteristic temperature dependence to the $O(4)$ criticality,
represented by $R_{6,2}$ which is negative around $T_{pc}$  and  approaches to zero at high
temperature,  is lost. Instead, $R_{6,2}$ exhibits strong suppression to
a larger negative values, due to the ultraviolet momentum cutoff.

\begin{figure}[ht]
 \centering
 \includegraphics[width=0.6\textwidth]{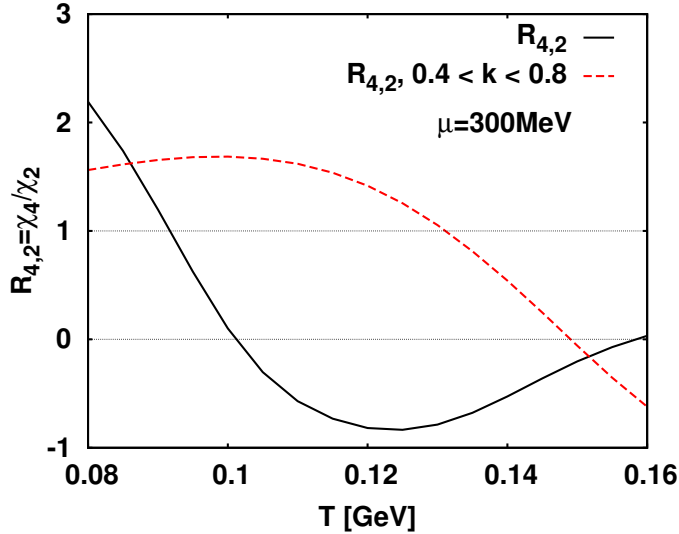}
 \caption{As in Fig.~\ref{fig:cum62_cutoff}-right but the calculations
 are done at fixed chemical potential $\mu=300$ MeV.  }
  \label{fig:c4c2_finitemu_cutoff}\vskip0.3cm
\end{figure}

The effects of  momentum cuts are even stronger at finite chemical
potential. Fig.~\ref{fig:c4c2_finitemu_cutoff} shows the $R_{4,2}$ as a
function of temperature  at $\mu=300$MeV  where $T_{pc}=125$ MeV.
Contrary to the case of  $\mu=0$, the cutoff changes
the sign of $R_{4.2}$  near the chiral cross over. While the negative structure  of $R_{4.2}$ at
$T_{pc}=125$ MeV signals the remnant of the $O(4)$ criticality,
 the infrared cutoff $k > 2.2m_\pi$ implies a changing  the  sign of
 $R_{4.2}$.

 Figures~\ref{fig:cum62_cutoff} and \ref{fig:c4c2_finitemu_cutoff} make
 it clear,   that imposing cutoffs in momentum scale, modifies  the
 characteristic property of the cumulants ratio, which are specific to
 the  chiral crossover  at finite and vanishing chemical potential near
 the $O(4)$ pseudocritical points. 

\section{Concluding remarks}\label{sec:summary}

We have studied the momentum scale dependence of the net baryon number
fluctuations near chiral crossover, which appear in the critical region
of the second order phase transition in the $O(4)$ universality class. Our
calculations were done in  the quark-meson model within the functional
renormalization group (FRG) approach at finite and vanishing chemical
potential.
We have applied  the momentum scale dependence of the FRG flow equation
to quantify, how the suppression of the momentum modes in the infrared
and ultraviolet regimes modifies generic properties of the net baryon
number fluctuations ratio,  expected from remnants of the $O(4)$
criticality.

We have shown, that the  pion mass $m_\pi$ is a natural infrared soft
momentum scale at which cumulants are saturated at their critical
values, whereas  for  scales larger than  $2m_\pi$  the characteristic
$O(4)$ structure of the higher order cumulants get lost.

In the ultraviolet regime, the momentum cutoff implies suppression of
different   cumulants ratios $R_{n,m}$. This suppression  is small for
$R_{n,m}$ which are insensitive to the chiral criticality. However, it
is essential for ratios which are directly linked to the singular part
of the free energy density, that is  responsible for remnants of the
$O(4)$ criticality in physical observables.

 The above properties of different net baryon number cumulant ratios
 in a model with the $O(4)$  chiral critical behavior are in
contrast to  models with the Skellam probability distribution of the net
 baryon number, which is   used as a reference for a non critical
 behavior.    Imposing any momentum cutoffs in such models with Skellam
 function changes the values of different cumulants but preserves
 their ratios. These results indicate, that when measuring
 fluctuations of the net baryon number in heavy ion collisions to search
 for a  restoration of a chiral symmetry or critical point, a
 special care has to be made when introducing kinematical cuts on the
 fluctuation measurements. While there is no direct relation
 between the kinematical cuts imposed on measured particle momenta  and the momentum scale cut
 in the flow equation \eqref{eq:floweq}, one expects that the low $p_T$
 particles are more affected by the soft modes in a medium. One also
 finds explicitly, that the scale momentum $k$ in the flow equation
 \eqref{eq:floweq} reduces to the particle momentum in the case of a
 free gas of  quarks and mesons.
 In turn, this also implies,  that one should
 observe modifications of the  higher order cumulants ratios against the
 variation of the momentum cutoff,  if they  are influenced by the chiral
 critical behavior or its remnant. 

\section*{Acknowledgment}

We acknowledge  stimulating discussions with Bengt Friman and Chihiro Sasaki.
This work was supported by the Polish Science Foundation (NCN), under
Maestro grant 2013/10/A/ST2/00106.
K.M. was  supported by the Grant-in-Aid for Scientific Research on
Innovative Areas from MEXT (No. 24105008).


%

\end{document}